\documentclass[sigconf]{acmart}
\usepackage{multirow} 
\usepackage[utf8]{inputenc}

\AtBeginDocument{%
  }

\acmConference[BioKDD '24]{BIOKDD}{August 25,
  2024}{Barcelona, Spain}

\begin{document}

\title{Pre-Trained Foundation Model representations to uncover Breathing patterns in Speech}

\author{Vikramjit Mitra}
\email{vmitra@apple.com}
\orcid{0000-0002-2721-3976}
\affiliation{%
  \institution{Apple}
  \city{Cupertino}
  \state{California}
  \country{USA}
}

\author{Anirban Chatterjee}
\email{anirbanc@apple.com}
\affiliation{%
  \institution{Apple}
  \city{Cupertino}
  \country{USA}}

\author{Ke Zhai}
\email{ke_zhai@apple.com}
\affiliation{%
  \institution{Apple}
  \city{Cupertino}
  \country{USA}
}

\author{Helen Weng}
\email{helen_weng@apple.com}
\affiliation{%
  \institution{Apple}
  \city{Cupertino}
  \country{USA}
}

\author{Ayuko Hill}
\email{amorikawa@apple.com}
\affiliation{%
  \institution{Apple}
  \city{Cupertino}
  \country{USA}
}

\author{Nicole Hay}
\email{nicole_hay@apple.com}
\affiliation{%
  \institution{Apple}
  \city{Cupertino}
  \country{USA}
}

\author{Christopher Webb}
\email{sudo@apple.com}
\affiliation{%
  \institution{Apple}
  \city{Cupertino}
  \country{USA}
}

\author{Jamie Cheng}
\email{jamie_cheng@apple.com}
\affiliation{%
  \institution{Apple}
  \city{Cupertino}
  \country{USA}
}

\author{Erdrin Azemi}
\email{erdrin@apple.com}
\affiliation{%
  \institution{Apple}
  \city{Cupertino}
  \country{USA}
}
\renewcommand{\shortauthors}{Mitra et al.}

\begin{abstract}
  The process of human speech production involves coordinated respiratory action to elicit acoustic speech signals. Typically, speech is produced when air is forced from the lungs and is modulated by the vocal tract, where such actions are interspersed by moments of breathing in air (inhalation) to refill the lungs again. Respiratory rate ($RR$) is a vital metric that is used to assess the overall health, fitness, and general well-being of an individual. Existing approaches to measure $RR$ (number of breaths one takes in a minute) are performed using specialized equipment or training. Studies have demonstrated that machine learning algorithms can be used to estimate $RR$ using bio-sensor signals as input. Speech-based estimation of $RR$ can offer an effective approach to measure the vital metric without requiring any specialized equipment or sensors. This work investigates a machine learning based approach to estimate $RR$ from speech segments obtained from subjects speaking to a close-talking microphone device. Data were collected from N=26 individuals, where the groundtruth $RR$ was obtained through commercial grade chest-belts and then manually corrected for any errors. A convolutional long-short term memory network (\textbf{\texttt{Conv-LSTM}}) is proposed to estimate respiration time-series data from the speech signal. We demonstrate that the use of pre-trained representations obtained from a foundation model, such as Wav2Vec2, can be used to estimate respiration-time-series with low root-mean-squared error and high correlation coefficient, when compared with the baseline. The model-driven time series can be used to estimate $RR$ with a low mean absolute error ($MAE$) ${\approx 1.6 \, breaths/min}$. 
\end{abstract}



\received{5 June 2024}
\received[revised]{13 July 2024}
\received[accepted]{1 July 2024}

\maketitle

\section{Introduction}
The lungs play a central role in speech vocalization, where they act as the source of air that is pumped through the vocal tract, which acts as a filter \cite{stevens2000acoustic} to generate acoustic speech. Breathing is the source of most sounds that humans vocalize and speech production requires control and coordination of breathing and speech articulation, also known as speech breathing \cite{fuchs2021respiratory}. Speech breathing demands more effort than regular breathing, where speech breathing is characterized by short inhalations to minimize interruptions during speech production, whereas regular breathing consists of equal phases of inhalation and exhalation \cite{hixon1987respiratory}. Due to short inhalations, the velocity of air-inflow is higher compared to regular breathing \cite{conrad1979speech}, hence, breath sound is normally audible in speech \cite{routray2019automatic}. The volume of air exhaled during speech is influenced by the length and loudness of the intended utterance, and the exhale-duration is dependent upon the linguistic intent and sounds produced during speech production \cite{winkworth1994variability, klatt1968studies}. Speech production and breathing are inherently coupled and \cite{nallanthighal2020speech} aimed at sensing speech breathing patterns from the linguistic content and prosodic factors of speech.

Respiratory rate ($RR$) is a vital metric, where studies have shown that $RR$ is the most valid marker of exertion \cite{nikolo2014a, nikolo2017} and a reduction in $RR$ is an indicator of a person's relaxation response \cite{grant, wielgosz, kral, kral2023slower} and self-reported well-being \cite{kral2023slower}. Speech breathing parameters have been used for clinical applications \cite{solomon1993speech} as well as for affective analysis \cite{goldman1955speech, heim1968emotion}. Prior work on breath-sound detection from audio has focused on the detection and categorization of particular breath sounds to distinguish between healthy and abnormal breath sounds \cite{li2017design, castro2014real}. $RR$ estimation has been investigated from both contact-based sensors and non-contact-based sensors \cite{sierra2006comparison, sierra2004monitoring, ren2015fine, kumar2021estimating, ahmed2023remote, rahman2022breathebuddy}, to acquire nasal breath recordings and wearable microphones. In this work, we investigate estimating respiratory parameters from speech recorded using close-talking microphones, that is more likely to sense respiratory sounds in speech, compared to distant-microphones, due to their proximity to the mouth. 

Prior work on speech-breathing focused mostly on using traditional acoustic features such as log-mel spectrograms \cite{nallanthighal2020speech}, or their discrete cosine transformed counterparts (a.k.a, mel-frequency cepstral coefficients or MFCCs) \cite{routray2019automatic, ruinskiy2007effective, macintyre2020deep}. However, in case of limited-size datasets, such representations make the downstream machine learning models prone to over-fitting, and as a consequence restrict the generalization capacity and robustness of the machine learning (ML) model. Recent advances in foundation models \cite{bommasani2021opportunities} have resulted in significant performance boost of speech technologies, where pre-trained model representations \cite{baevski2020wav2vec, hsu2021hubert} have shown state-of-the-art performance for speech recognition \cite{zuluaga2023does}, speaker recognition \cite{zuluaga2023does}, and emotion recognition \cite{mitra2022speech}. Representations from pre-trained foundation models have demonstrated better generalization capacity and robustness across different speech tasks, under various acoustic conditions and for multiple languages, hence we hypothesize that such representations will be quite useful for the task of speech based respiration parameter estimation.

Self-supervised learned (SSL) models such as Wav2Vec2 \cite{baevski2020wav2vec} or HuBERT \cite{hsu2021hubert} are trained on large volumes of unlabeled data and are anticipated to learn acoustic units from the training data. The learned acoustic units should be discriminable in their spectro-temporal representations, and represent distinct acoustic phonetic units (such as vowels, voiced/unvoiced consonants, pauses, aspirated noise etc.) or their sub-states. 

\vspace{1mm}

\noindent In this work, we aim to:
\newline (1) estimate the respiration time-series signal from speech data, 
\newline (2) obtain $RR$ measure from speech data, and 
\newline (3) detect inhale events within the speech data. 
\vspace{1mm}

 We hypothesize that pre-trained representations should have information that can help with the above tasks and demonstrate better performance compared to standard mel-filterbank (MFB) based acoustic features given that they are pre-trained with large speech datasets. 

\vspace{1mm}
\noindent This work demonstrates that:
\newline (1) features from pre-trained models significantly improve $RR$ estimation from speech compared to standard acoustic features.
\newline (2) respiration time-series (inhale/exhale signal) can be estimated from speech using an ML model, that is highly correlated to the reference measures.
\newline (3) saliency-driven pre-trained representations can reduce the dimensionality of input representation space, as a consequence can reduce the downstream model's parameter size. 
\newline (4) fusing pre-trained representations with standard acoustic features can improve $RR$ estimation performance.

\vspace{1mm}
Note that unlike prior works \cite{nallanthighal2020speech, routray2019automatic} that have used standard acoustic features, we demonstrate that pre-trained model representations can be used for speech breath detection, and can demonstrate superior performance. In addition, we present a metric (breath-event error rate: $BER$) that indicates how closely the detected breath-events align with the groundtruth data. Finally, we present a convolution LSTM (\textbf{\texttt{Conv-LSTM}}) model and show that the network-depth and fusion of pre-trained representations and MFB helps to better estimate the breath time-series data from speech.

The rest of the paper is organized as follows: Section (2) presents the dataset used in our study, Section (3) introduces feature representations investigated and details on the acoustic model and its parameters, Section (4) presents the results, followed by conclusions in Section (5).

\section{Data}
\label{sec:Data}
Publicly available speech datasets containing respiration time-series reference do not exist, hence we collected data internally. The 2020 speech paralinguistic challenge \cite{schuller2020interspeech} explored speech based respiration event detection, however the dataset used in that challenge is not publicly available. Data were collected from 26 adult speakers under realistic background acoustic environments (consisting of background noise) in an indoor setting. American English speakers, between the age 25 to 60, balanced by gender, were employed for the data collection. Data were recorded using microphone-enabled, wearable headphones. Speech data collected using wearable microphones and chest-belt measurements were collected across multiple sessions. During the data collection, participants were prompted to read a paragraph, where the reading session varied from 45 to 90 seconds. Note that conversational speech is not considered in this study, however we expect that findings from this work should generalize to such speech. 

\begin{figure*}[h]
  \centering
  \includegraphics[width=0.85\textwidth]{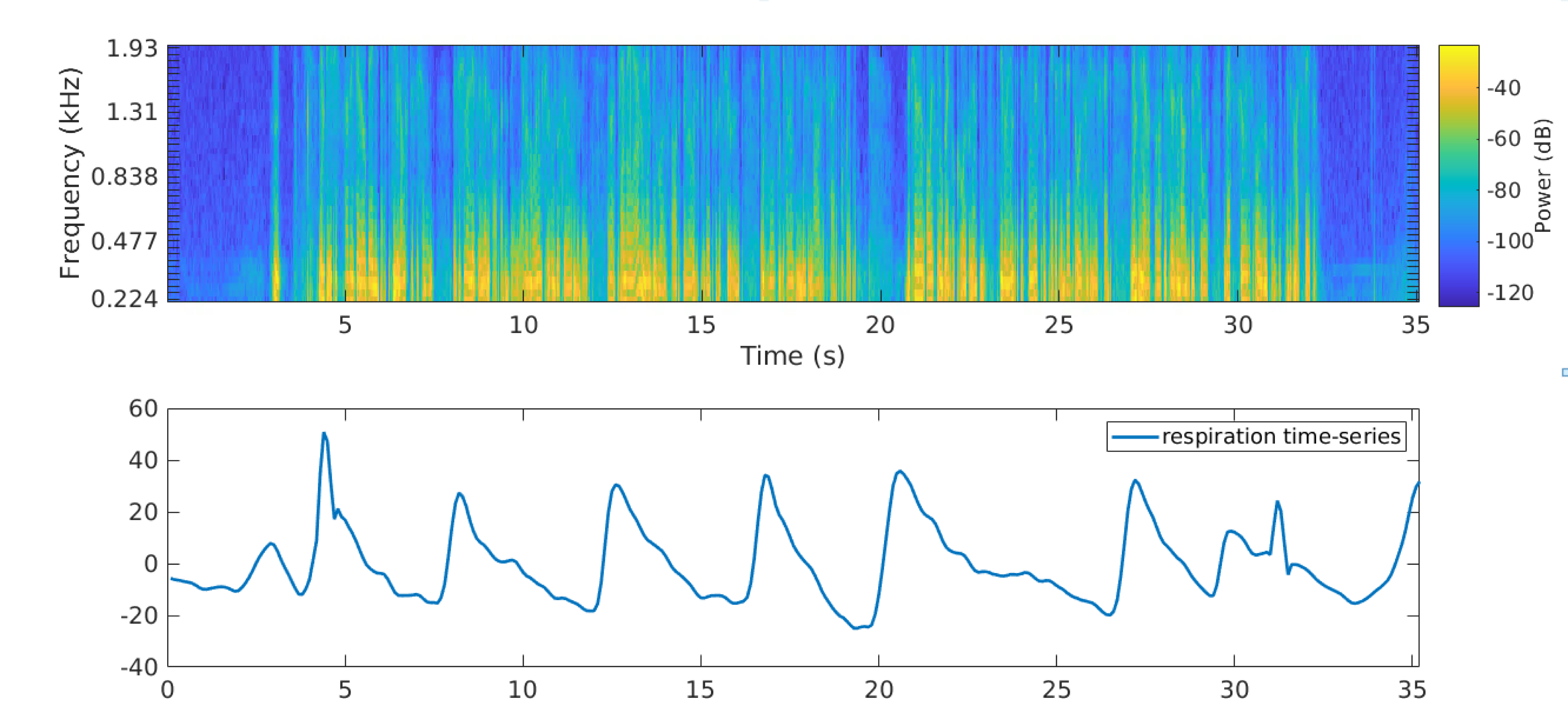}
  \caption{Spectrogram speech [top] and the corresponding chest-belt pressure measurement (in Newton) [bottom].}
  \label{fig:fig1}
  \raggedleft
\end{figure*}

A strain-gauge chest-belt sensor (Vernier Go Direct Respiration Belt) was used during the data collection to obtain groundtruth reference chest contraction and relaxation (corresponding to inhalation and exhalation) measurements. Figure \ref{fig:fig1} shows a plot of a sample respiration signal spectrogram and its corresponding chest belt measurement. Due to calibration and subject variability, chest-belt measurements were observed to have variations, hence a quality check of the chest-belt measurement was performed manually and any data with erroneous measurement were removed. Chest-belt data were z-score normalized and dynamic range compressed before being used for model training. For some sessions the participant did not speak, hence they did not contain any recorded speech; such data were excluded from our experiments. 

Data augmentation was performed to simulate faster and slower breathing by altering the speed of the entire audio signal. We have used 25 hours of speech data from 26 speakers in this study, where data from 3 and 4 speakers (roughly one hour of speech data/speaker) were set aside for validation and test sets, and the remaining 19 speakers were used for model training. Note that the validation and test split speakers were balanced by gender. Speech data were segmented into chunks of 30 seconds for model training, to ensure it contains at least one full breath cycle.

\setlength{\textfloatsep}{2pt plus 1.0pt minus 1.0pt}

\setlength{\textfloatsep}{2pt plus 1.0pt minus 1.0pt}
\begin{figure}[H]
  \centering
  \includegraphics[width=0.5\textwidth]{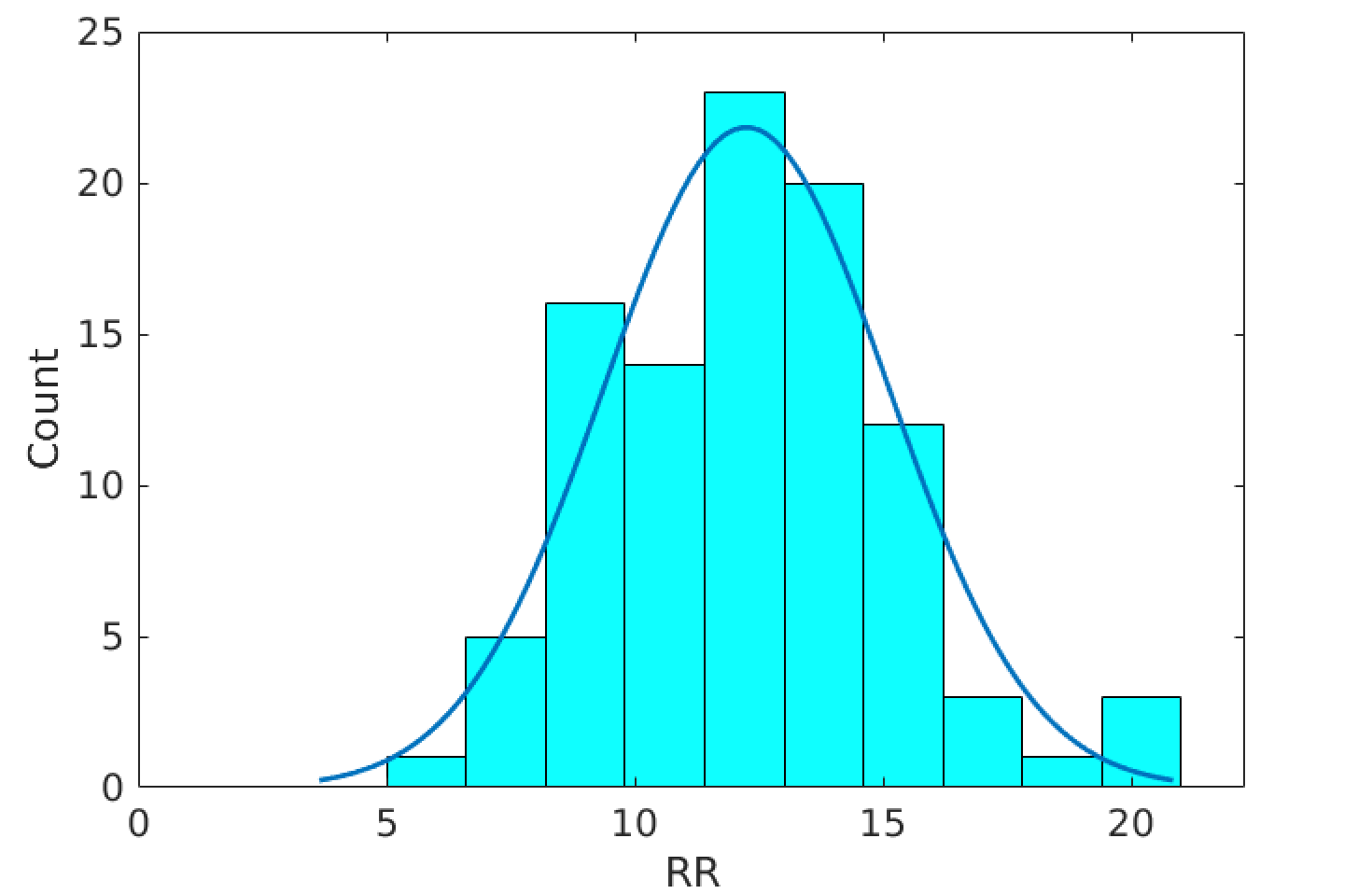}
  \caption{Histogram of $RR$ (in br/min) estimated from chest-belt data in the dataset.}
  \label{fig:fig2}
  \raggedleft
\end{figure}

\subsection{Analysis} 
\label{sec:Analysis} 
\begin{figure}[H]
  \centering
  \includegraphics[width=0.47\textwidth]{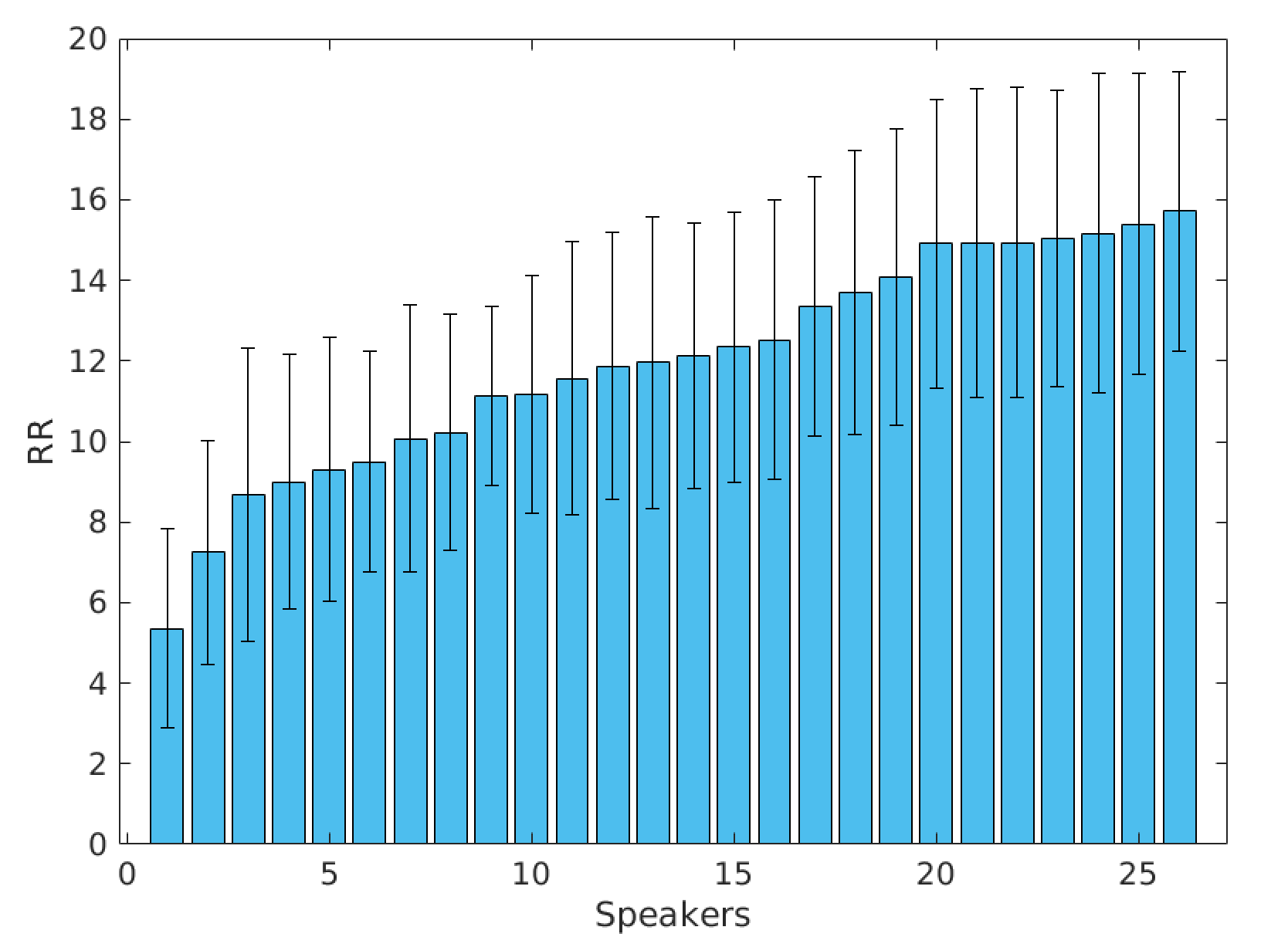}
  \caption{Mean and std-dev $RR$ (in br/min) by speakers.}
  \label{fig:fig3}
  \raggedleft
\end{figure}

We analyzed the data used in this study to measure the variance in $RR$, both within and across speakers. Figure \ref{fig:fig2} shows the histogram of $RR$ estimated from the chest-belt data obtained from the subjects in our dataset. Figure \ref{fig:fig3} shows the variance of $RR$ by subject, which shows that $RR$ varied not only across subjects, but also within the subject across multiple sessions. The overall dynamic range of $RR$ values in the dataset were within the range of 5 to 19 breaths/min (denoted as br/min).

\section{Methods}
\subsection{Acoustic Features}
The baseline acoustic features consist of 40-dimensional MFB energies, analyzed at a 25ms window, with a frame interval of 10ms. 

\setlength{\textfloatsep}{2pt plus 1.0pt minus 1.0pt}

\subsection{Features from Pre-Trained Models}
We explored embeddings generated from a pre-trained Wav2Vec2-base (Wav2Vec2) model \cite{baevski2020wav2vec}\footnote{we have selected Wav2Vec2-base due to its smaller size}. Note that the pre-trained acoustic model was not fine-tuned to our data, and its parameters were frozen to generate the representations for our dataset. The Wav2Vec2 model was pre-trained on 960 hours of speech from the \textit{Librispeech} dataset with 12 transformer layers and 768 embedding dimensions, where we investigated the representations obtained from the $2^{nd}$ through the last transformer layers\footnote{\url{https://pytorch.org/audio/stable/pipelines}}. Representations from the initial layers are expected to contain more acoustic information, while those from the latter layers are expected to contain more phonetic information.

\begin{figure*}
\begin{minipage}[b]{1.0\linewidth}
  \centering
  \centerline{\includegraphics[width=0.95\textwidth]{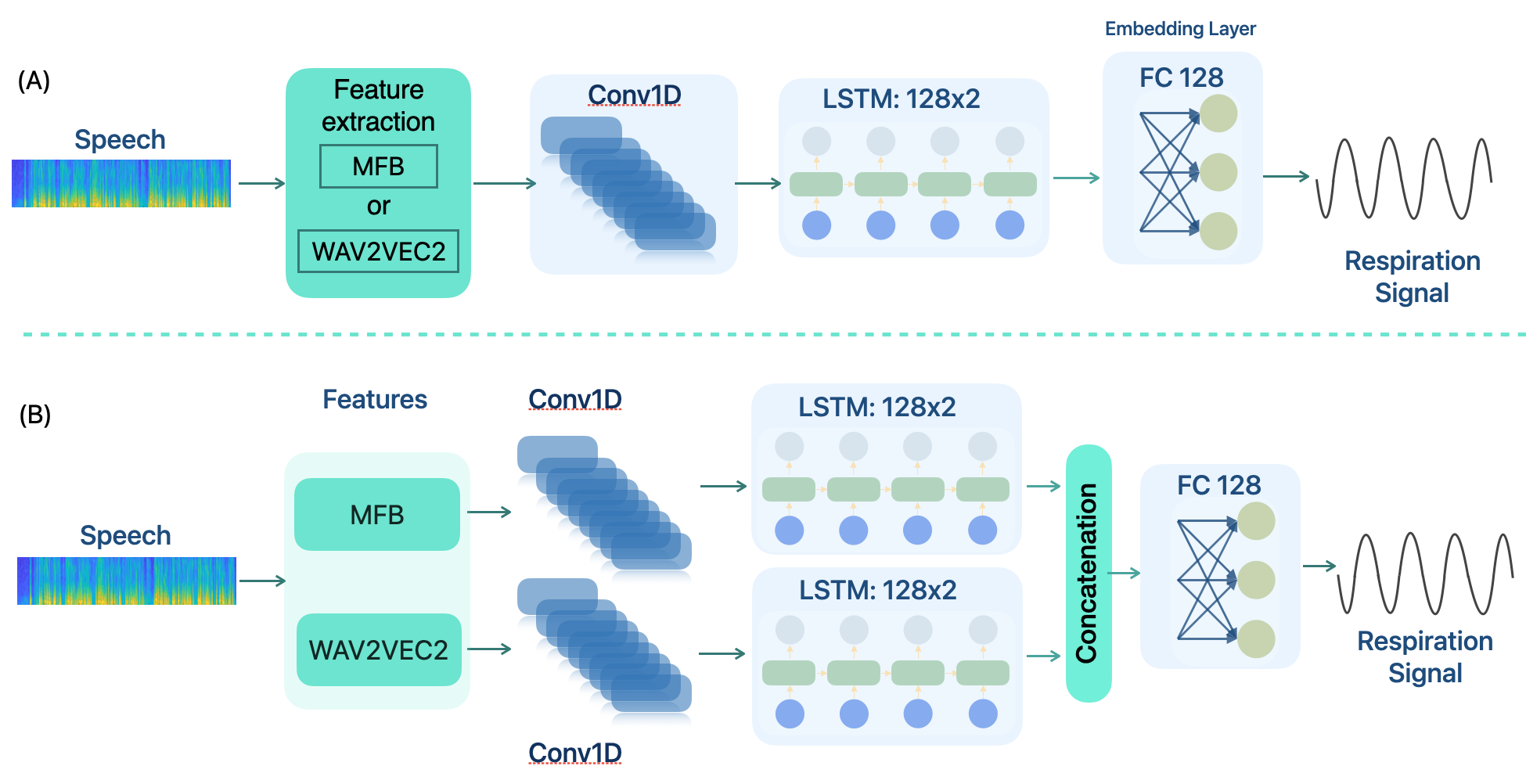}}
\end{minipage}
\caption{(A) Architecture of the single-feature (\textbf{\texttt{Conv-LSTM}}) network, and (B) Feature-fused network} 
\label{fig:fig4}
\end{figure*}

\subsection{Model}
We used a convolutional network with Long-Short term memory units (\textbf{\texttt{Conv-LSTM}}) consisting of as many time-convolution filters as the number of feature inputs (which is 40 for MFB and 768 for Wav2Vec2), 128 LSTM units and 128 neurons in the embedding layer. The model architecture is shown in Figure \ref{fig:fig4}.A. Additionally, we investigated feature fusion as shown in Figure \ref{fig:fig4}.B. Given the ability of foundation models (such as Wav2Vec2) to learn large dimensional acoustic representations through multiple tiers of transformer layers, the down-stream classifiers trained on the foundation model representations can be simple in architecture, as reported in \cite{mitra2022speech}. In this work we did not observed any evidence of performance gain by increasing model complexity (by introducing additional layers), hence we focused on exploring a simple (\textbf{\texttt{Conv-LSTM}}) architecture as shown in Figure \ref{fig:fig4}.A.

Models were trained using the concordance correlation coefficient ($CCC$) \cite{lawrence1989concordance} as the loss function (see Equation (1)). In Equation (1), where ${\mu _{x}}$ and ${\mu _{y}}$ are the means, ${\sigma _{x}^{2}}$ and ${\sigma _{y}^{2}}$ are the corresponding variances for the estimated and groundtruth time-series data, and ${\rho}$ is the correlation coefficient between the two variables. The models were trained with a mini-batch size of 64, using Adam optimizer with a learning rate of 0.005. Early stopping was performed based on the validation-set loss. 

\begin{equation}
\begin{aligned}
CCC &= \frac {2\rho \sigma_x \sigma_y}{\sigma_x^2+\sigma_y^2 +(\mu_x-\mu_y)^2 }.
\end{aligned}
 \label{eq1}
\end{equation}

\subsection{Salient representations}
\label{sec:salient}
The pre-trained model embeddings have large dimensionality, for example, Wav2Vec2 model generates 768 dimensions, resulting in increased downstream model size. To reduce the feature dimension, we obtained breath-salient representations from the Wav2Vec2, by relying on the relationships between the input representation and the targets. Prior studies \cite{mitra2020investigation, mitra2023investigating} have explored the input-output relationships of activations to obtain neural saliency, and we use a similar idea to obtain salient representations for respiration signal estimation. Let the $k^{th}$ dimension of $N$ dimensional Wav2Vec2 for an utterance $y$ be represented by a vector $H_{k,y} = [X_{1,k}, \dots, X_{M,k}]$, where $M$ denotes the sequence length. Let the reference respiration time-series be $L$ for utterance $y$. The cross-correlation based saliency ($CCS_k$) of $k^{th}$ dimension is given by:

\newcommand\normx[1]{\left\Vert#1\right\Vert}
\begin{equation}
S_{CCS_k} = \normx {\frac {Cov({{H}_k},L)}{\sigma_{H_k}\sigma_L}} + \gamma_k,
    \label{eq2}
\end{equation}

where Equation \ref{eq2} computes the absolute cross-correlation between time-series $L$ and embeddings ${{H}_{k}}$ for dimension $k$ for all utterances in the \textit{training} set. $\gamma_k$ is the sum of the weighted cross-correlation between the $k^{th}$ dimension and all other dimensions, as shown in Equation \ref{eq3}:

\begin{equation}
\gamma_k = \frac {1}{N-1} \sum_{j=1, j \ne k}^{N} w_j \normx {\frac {Cov({{H}_k},{{H}_j})}{\sigma_{{{H}_k}}\sigma_{{{H}_j}}}}, \\ 
\label{eq3}
\end{equation}

\begin{center}
where, $w_j = \normx {\frac {Cov({{H}_j},L)}{\sigma_{{{H}_j}}\sigma_L}}$.
\end{center}

In our experiments we have used $S_{CCS}$ given in Equation \ref{eq2} to select salient dimensions in pre-trained representations.

\begin{table*}
  \caption{Baseline performance (on test-set) for respiration time-series estimation using $CCC$ and $RMSE$ measures using MFB and Wav2Vec2 representations}
  \label{tab:table1}
  \begin{tabular}{lccccc}
    \toprule
     \multirow{2}{*}{\textbf{Representations}} & \multirow{2}{*}{\textbf{Layer}} &
       \multicolumn{2}{c}{\textbf{Time Series}} &
       \multicolumn{2}{c}{\textbf{RR Estimate}} \\
         & & {$CCC \uparrow$} & {$RMSE \downarrow$} & {$MAE \downarrow$} & {$Acc@2bpm \uparrow$} \\
    \midrule
     MFB & N/A & 0.68 & 0.13 & 2.85 & 64.1\\
       \midrule
      & 2 & 0.73 & 0.12 & 2.67 & 66.4\\
      & 3 & 0.75 & 0.12 & 2.56 & \textbf{67.2}\\
      & \textbf{4} & \textbf{0.76} & \textbf{0.11} & 2.52 & 66.2\\
      & 5 & 0.73 & 0.12 & 2.59 & 66.0\\
      & 6 & 0.75 & 0.12 & 2.67 & 66.3 \\
      $Wav2Vec2$ & \textbf{7} & \textbf{0.76} & \textbf{0.11} & 2.56 & 65.5\\
      & 8 & 0.75 & 0.12 & \textbf{2.35} & 66.7\\
      & 9 & 0.75 & 0.13 & 2.56 & 64.8\\
      & 10 & 0.74 & 0.13 & 2.57 & 64.6\\
      & 11 & 0.71 & 0.12 & 2.75 & 63.8\\
      & 12 & 0.69 & 0.12 & 2.86 & 63.5\\
    \bottomrule
  \end{tabular}
\end{table*}

\begin{table*}
  \caption{Performance on Validation set for respiration time-series estimation using $CCC$ and $RMSE$ measures using MFB and Wav2Vec2 representations}
  \label{tab:table2}
  \begin{tabular}{lccccc}
    \toprule
     \multirow{2}{*}{\textbf{Representations}} & \multirow{2}{*}{\textbf{Layer}} &
       \multicolumn{2}{c}{\textbf{Time Series}} &
       \multicolumn{2}{c}{\textbf{RR Estimate}} \\
         & & {$CCC \uparrow$} & {$RMSE \downarrow$} & {$MAE \downarrow$} & {$Acc@2bpm \uparrow$} \\
    \midrule
     MFB & N/A & 0.57 & 0.15 & 3.61 & 61.5\\
       \midrule
      & 2 & 0.62 & 0.14 & 2.89 & 62.7\\
      & 3 & 0.63 & 0.14 & 2.64 & 64.7\\
      & 4 & 0.66 & 0.14 & 2.32 & 67.8\\
      & 5 & 0.65 & 0.14 & 2.54 & 68.1\\
      & 6 & 0.66 & 0.14 & 2.60 & 66.2 \\
      $Wav2Vec2$ & 7 & 0.67 & 0.13 & 2.35 & 67.4\\
      & 8 & 0.66 & 0.14 & 2.43 & 69.3\\
      & 9 & 0.63 & 0.14 & 2.55 & 66.7\\
      & 10 & 0.59 & 0.14 & 2.57 & 65.6\\
      & 11 & 0.59 & 0.15 & 2.71 & 63.3\\
      & 12 & 0.58 & 0.15 & 2.91 & 62.2\\
    \bottomrule
  \end{tabular}
\end{table*}

\begin{figure*}
\begin{minipage}[b]{1.0\linewidth}
  \centering
  \centerline{\includegraphics[width=0.8\textwidth]{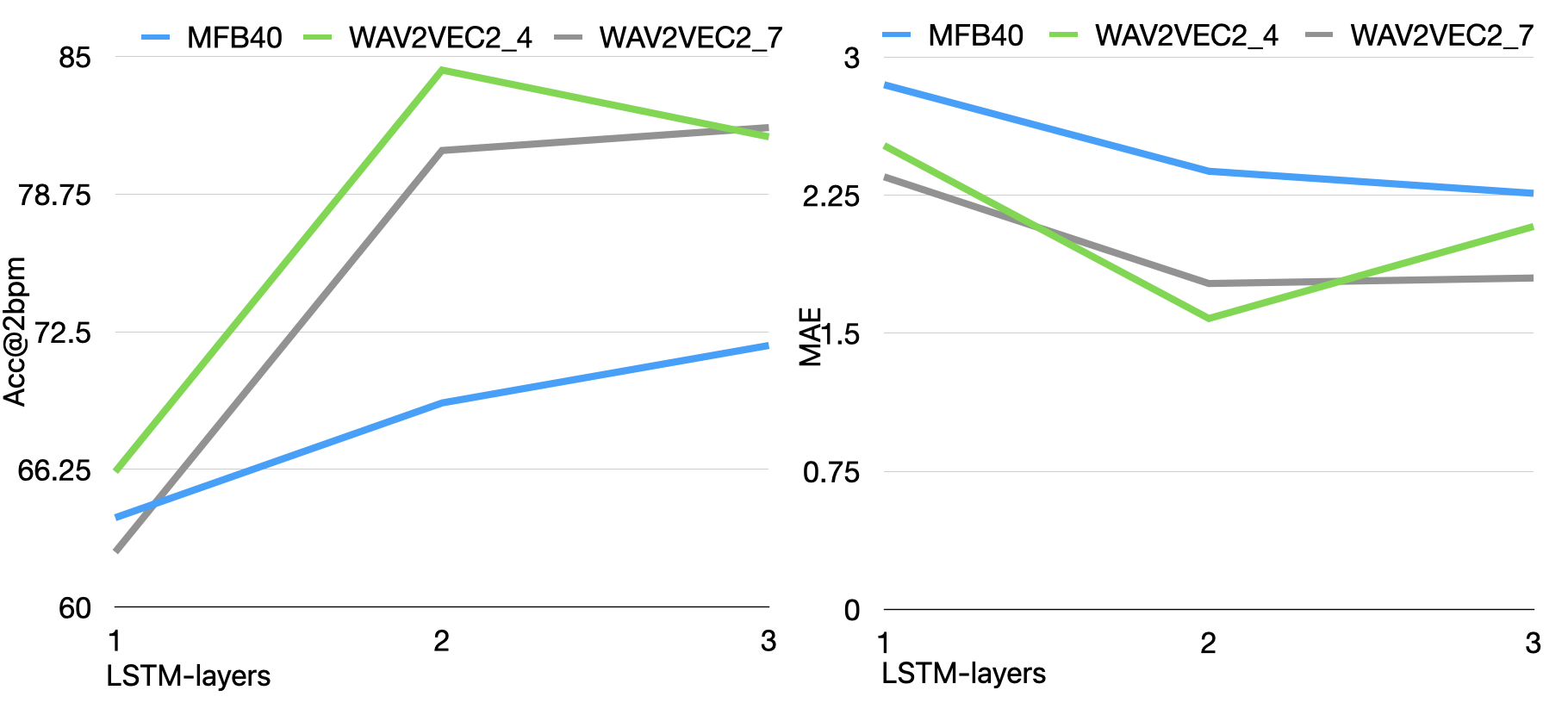}}
\end{minipage}
\caption{Segment-level performance by number of LSTM layers for models trained with MFB and Wav2Vec2 features} 
\label{fig:fig5}
\end{figure*}

\begin{figure*}
  \centering
  \includegraphics[width=0.85\textwidth]{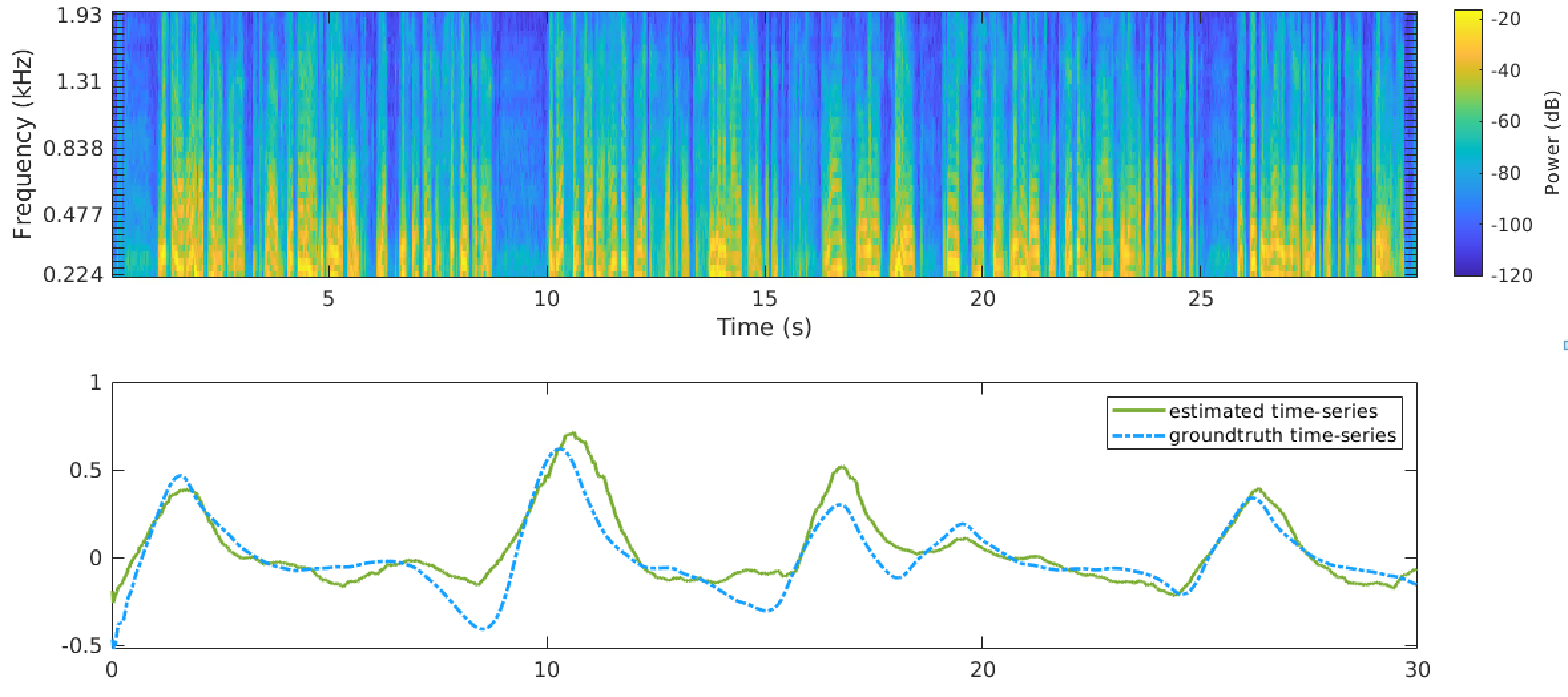}
  \caption{Spectrogram speech [top] and the corresponding chest-belt time-series (grountruth) in blue and the estimated time-series from the model in green [bottom].}
  \label{fig:fig6}
  \raggedleft
\end{figure*}

\section{Results}

We trained baseline acoustic models using (i) MFB and (ii) Wav2Vec2 embeddings obtained from the $2^{nd}$ through $12^{th}$ transformer layers of the model. The performance of the respiration time-series estimation model is shown in Table \ref{tab:table1}. We present the results using metrics focusing on the time-series respiration signal estimation, where we have used $CCC$ \cite{lawrence1989concordance} and root-mean-squared error ($RMSE$). Table \ref{tab:table1} shows the baseline time-series estimation performance obtained from MFB and Wav2Vec2 representations. We also evaluated the segment-level $RR$ estimation performance, where for segment-level $RR$ estimation, we have used the following metrics: mean-absolute error ($MAE$) and Accuracy at 2 br/min error tolerance ($Acc@2bpm$). $MAE$ is computed by comparing the number of breath-events detected from the estimated time-signal obtained from the model, with that observed in the chest-belt groundtruth signal.

Accuracy for a segment is measured at a tolerance bound of +/-2 breaths/min (bpm) (we made this selection to have a conservative error-bound), where an estimate outside the tolerance-bound is treated as an error. Table \ref{tab:table1} shows that the pre-trained representations from Wav2Vec2 perform better than the MFB features for the test-set, and the relative improvement was at-least 2.4\% increase in $CCC$ and  6.8\% relative reduction in $RMSE$. 

Interestingly, Table \ref{tab:table1} also shows that representations from different transformer layers of the Wav2Vec2 features had different impact on the performance, where the representations from layers 4 to 9 were more effective than the final layers 10 through 12. The best performance was obtained from layers 4 and 7, which gave 12.3\% relative improvement in $CCC$, and 14.3\% relative reduction in $RMSE$ compared to the MFB features. Even though we have used the SSL trained Wav2Vec2 (which is not fine-tuned on any specific task), the final layers may contain more phonetic-discriminatory information which may not be essential for breath-signal estimation (see section 3.2). The middle layers may contain more broad acoustic-level information that helps to detect the breathing patterns in speech, speech-activity and silent pauses, hence, they helped to generate better performance than the final layers. Note that given the findings in Table \ref{tab:table1}, we will be using the representations from layers 4 and 7 in the remaining of this paper to train (\textbf{\texttt{Conv-LSTM}}) models with 2 LSTM layers. 

Next, we investigated the depth of the LSTM layers and Figure \ref{fig:fig5} shows that a 2-layered LSTM model overall performed the best providing higher $Acc@2bpm$ and lower $MAE$ for all the features. Table \ref{tab:table2} show the validation set performance, when MFB feature and representations from different transformer layers of Wav2Vec2 was used.

We also investigated if saliency-driven feature selection can help to reduce the model size, while retaining the model performance. Using the approach outlined in section \ref{sec:salient} we investigated pruning input representations, by keeping only 90\%, 75\%, 50\% and 25\% of the input representations, which in turn resulted in reducing the model parameter size by 9\%, 22\%, 44\% and 66\% respectively. Table \ref{tab:table3} shows the result obtained from selecting salient representations from Wav2Vec2 layers 4 and 7. We introduce a metric: breath error rate ($BER$) to measure the accuracy of detecting breath events. $BER$ is computed by comparing the inhalation events in the groundtruth and estimated time-series signals, where we have only deletion of inhale-events (deletion errors, $D$) and inserted inhale-events (insertion errors, $I$), and use the total number of inhale events $N$ in the groundtruth data, to measure $BER$:

\begin{equation}
BER = {\frac {I + D}{N}},
\label{eq4}
\end{equation}

\begin{table*}
  \caption{Respiration time-series estimation performance (in $CCC$ and $RMSE$) and segmental $RR$ estimation (in $MAE$, $Acc@2bpm$ and BER) from Wav2Vec2 layers 4 and 7 and fusion of layer 4 with MFB, after saliency based representation selection and their corresponding parameter size reduction}
  \label{tab:table3}
  \begin{tabular}{lcccccccc}
    \toprule
     \multirow{2}{*}{Feature} & 
       \multicolumn{1}{c}{\%Input} &
       \multicolumn{2}{c}{Time Series} &
       \multicolumn{3}{c}{$RR$ estimate} & \multicolumn{1}{c}{$\downarrow$ \% Rel.}\\
       & { Reps.} & {$CCC \uparrow$} & {$RMSE \downarrow$} & {$MAE \downarrow$} & {$Acc@2bpm \uparrow$} & {$BER \downarrow$} & {model size}\\
       \midrule
      & 100 & 0.75 & \textbf{0.11} & \textbf{1.58} & \textbf{84.4} & 29.8 & 0 \\
      & 90 & 0.76 & 0.12 & 1.89 & 77.6 & 26.8 & 8.8 \\
     $Wav2Vec2_4$ & 75 & 0.75 & \textbf{0.11} & 2.13 & 75.5 & 29.3 & 22.0 \\
      & 50 & 0.76 & \textbf{0.11} & 1.80 & 78.1 & 24.9 & 44.0 \\
      & 10 & 0.72 & 0.12 & 1.97 & 74.5 & 32.4 & 66.0 \\
     \midrule
      & 100 & \textbf{0.77} & \textbf{0.11} & 1.77 & 80.7 & 28.7 & 0 \\
      & 90 & \textbf{0.77} & \textbf{0.11} & 1.89 & 79.7 & 30.1 & 8.8 \\
     $Wav2Vec2_7$ & 75 & 0.76 & \textbf{0.11} & 2.12 & 74.0 & 29.1 & 22.0 \\
      & 50 & 0.76 & \textbf{0.11} & 1.91 & 76.6 & 28.3 & 44.0 \\
      & 10 & 0.72 & 0.12 & 2.21 & 72.4 & 37.4 & 66.0 \\
      \midrule 
      $Wav2Vec2_{4,50}$+MFB & 50 & \textbf{0.77} & \textbf{0.11} & \textbf{1.58} & 83.9 & \textbf{22.6} & 27.4 \\
     \bottomrule
  \end{tabular}
\end{table*}

Table \ref{tab:table3} shows that the representations from layer 4 performed better than those from layer 7, especially for the segment-level $RR$ metrics ($MAE$, $Acc@2bpm$ and $BER$). Selecting the top 50\% representation based on saliency resulted in the best $BER$ with some regression in $MAE$ and $Acc@2bpm$ compared to the model trained with the full layer 4 representations. Note that the 50\% representation based model is smaller than the full-representation based model by 44\% (Figure \ref{fig:fig6} show the time-series estimate from the model). The above findings indicate that: (1) the earlier layers of Wav2Vec2 contain more respiration-relevant representation that resulted in better performance across multiple metrics, (2) $RR$ estimation $MAE$ as low as 1.6 bpm can be achieved using speech as input data, where an $RR$ estimation accuracy as high as 84\% can be obtained for a tolerance of +/-2 bpm, and (3) saliency-based representation can help to reduce the model size by 44\% that can provide better $BER$ but some regression in $RR$ estimation performance. Note that for segment-level $RR$ estimation the $MAE$ and  $Acc@2bpm$ obtained from MFB are 2.38 and 69.3\% respectively, indicating that Wav2Vec2 representations performed better than MFBs for the segment-level metrics as well. We investigated fusion of 50\% salient layer-4 representation with MFB features ($Wav2Vec2_{4,50}$+MFB), result shown in the last row of table \ref{tab:table3}, where we observed that fusion of information helped to achieve the best BER, with comparable MAE and $Acc@2bpm$ from the best single-feature system (Wav2Vec2 layer 4), with 27\% reduction in model parameter size. The fusion results indicate that the Wav2Vec2 and MFB representations may have complementary information, hence their fusion resulted in improved performance.

\section{Conclusion}

In this work we demonstrated that respiration signal can be estimated from speech data collected through close-talking microphones. Results from our work has shown a time-series estimation performance with $CCC$ as-high-as 0.77 and an $RMSE$ as-low-as 0.11, where the groundtruth respiration signal was z-score normalized. At the segment-level, we observed that $RR$ can be estimated with a $MAE$ of 1.6 bpm. We also observed that pre-trained model representations from Wav2Vec2 SSL model performed better than standard MFB feature, providing a relative $MAE$ reduction of 33.6\% and relative improvement in estimation $CCC$ by 10\%. Additionally, we observed that fusion of Wav2Vec2 and MFB features provided the best overall performance. 

Future studies should explore the use of representations from fine-tuned foundation models with speech data containing respiration-relevant information. Additionally, the impact of subjective variance and the models' generalization capacity should be investigated using a dataset containing larger number of subjects than what was available in the dataset used in this study. A limitation of this study is that it uses a dataset containing read speech, future work should investigate spontaneous speech for estimating respiration signal.

\bibliographystyle{ACM-Reference-Format}
\bibliography{bioKDD24}

\appendix

\end{document}